\documentclass[conference]{IEEEtran}
\IEEEoverridecommandlockouts
\usepackage{cite}
\usepackage{amsmath,amssymb,amsfonts}
\usepackage{algorithmic}
\usepackage{graphicx}
\usepackage{textcomp}
\usepackage{xcolor}
\usepackage{lineno}
\usepackage[bookmarks=false]{hyperref}
\usepackage{cleveref}
\usepackage{graphicx}
\usepackage{booktabs}
\usepackage{threeparttable}
\usepackage{array}
\usepackage{calc}
\usepackage{caption}
\usepackage{subcaption}
\usepackage{subcaption}
\crefname{figure}{figure}{figures}
\Crefname{figure}{Fig.}{Fig.}
\def\BibTeX{{\rm B\kern-.05em{\sc i\kern-.025em b}\kern-.08em
    T\kern-.1667em\lower.7ex\hbox{E}\kern-.125emX}}
\begin{document}

\title{Multi-Modal Large Models Based  Beam Prediction: An Example Empowered by DeepSeek}

\author{
    Yizhu Zhao\(^\text{1}\), Li Yu\(^\text{1}\), Lianzheng Shi\(^\text{1}\), Jianhua Zhang\(^\text{1}\), Guangyi Liu\(^\text{2}\) \\
    \(^\text{1}\)State Key Laboratory of Networking and Switching Technology, \\
    Beijing University of Posts and Telecommunications, Beijing, China \\
    \{zhaoyizhu, li.yu, shilianzheng, jhzhang\}@bupt.edu.cn \\
    \(^\text{2}\)China Mobile Research Institute, Beijing, China \\
    \{liuguangyi\}@chinamobile.com
}
\maketitle

\begin{abstract}
Beam prediction is an effective approach to reduce training overhead in massive multiple-input multiple-output (MIMO) systems. However, existing beam prediction models still exhibit limited generalization ability in diverse scenarios, which remains a critical challenge. In this paper, we propose MLM-BP, a beam prediction framework based on the multi-modal large model released by DeepSeek, with full consideration of multi-modal environmental information. Specifically, the distribution of scatterers that impact the optimal beam is captured by the sensing devices. Then positions are tokenized to generate text-based representations, and multi-view images are processed by an image encoder, which is fine-tuned with low-rank adaptation (LoRA), to extract environmental embeddings. Finally, these embeddings are fed into the large model, and an output projection module is designed to determine the optimal beam index. Simulation results show that MLM-BP achieves 98.1\textbf{\%} Top-1 accuracy on the simulation dataset. Additionally, it demonstrates few-shot generalization on a  real-world dataset, achieving 72.7\textbf{\%} Top-1 accuracy and 92.4\textbf{\%} Top-3 accuracy with only 30\textbf{\%} of the dataset, outperforming the existing small models by over 15\textbf{\%}.
\end{abstract}


\section{Introduction}
Beamforming and massive MIMO technologies are desired to overcome the notorious path loss problem of high frequency bands \cite{ref1}. 
By forming a narrow beam, the base station can concentrate its transmission energy along a desired direction, thus improving the strength of the received signal \cite{ref2}.
However, in dynamic scenarios like the Internet of Vehicles (IoV), traditional methods for obtaining the optimal beam, such as sweeping the beam with a pre-defined codebook, require frequent beam training due to rapid changes in the optimal beam direction. 
Therefore, a major challenge in vehicle-to-everything (V2X) communication is to efficiently perform beam selection while avoiding training overhead.

With the rapid development of sensing and AI technology, modern vehicles are equipped with considerable computational capabilities and a variety of auxiliary sensors. These sensors capture data such as location and images, which contain wireless environment information (WEI)\cite{ref3}. 
WEI can characterize the spatial features of the current communication environment when the environment changes dynamically \cite{ref4}. 
Therefore, by leveraging WEI acquired at a lower cost through sensors, AI models can dynamically and instantly extract the hidden environment knowledge and utilize the knowledge for beam prediction\cite{ref5}. 
The utilization of images and location to assist in  beam selection has attracted growing interest in recent years. 
The authors of \cite{ref6} examine position-aided beam selection with GPS data, assessing training overhead reduction. In \cite{ref7}, a vision-aided beam alignment framework is proposed, utilizing camera images and 3D object detection to infer optimal beams. The authors of \cite{ref8} propose a multi-modal approach combining position and visual data, achieving higher prediction accuracy than single-modal methods.

However, existing studies have yet to address the generalization challenges faced by models deployed in vehicles, which struggle to adapt to frequent environmental changes caused by vehicle mobility. As a result, these models often tend to memorize specific mappings due to their limited parameters. Moreover, the scarcity of labeled real-world datasets imposes additional challenges on model generalization. Large language models (LLMs), which have achieved remarkable success in the natural language processing domain, demonstrate strong generalization capabilities and hold promise for addressing these challenges. Despite some researchers having applied LLMs to beam prediction \cite{ref9}, the text-only models used in these approaches fail to effectively handle multi-modal environmental features and lack validation of few-shot generalization in unseen scenarios. Multi-modal large models (MLMs) have demonstrated powerful multi-modal understanding and generalization capabilities. Hence, how to apply MLMs to beam prediction and rigorously evaluate their generalization capability warrants further investigation.

To the best of our knowledge, MLMs have not yet been applied in beam prediction research to understand information from different modalities. In this paper, we first propose a novel framework that utilizes MLMs for  beam prediction. This framework integrates multi-modal data obtained from vehicle sensors, including multi-view images and position data. The fusion of these two modalities mitigates image sensitivity to lighting and GPS's lack of environmental details \cite{ref10}. After preprocessing the inputs, we make several adjustments to the MLM. Specifically, we apply LoRA \cite{ref11} for lightweight fine-tuning of the image encoder and freeze the large model parameters except for the normalization layers to save computational resources. In addition, we design an output projection module to translate the output of the large model into a probability distribution for the optimal beam index. Through a large-scale environment-channel dataset collected from the digital world and a real-world dataset, we evaluate and demonstrate the performance advantages of the proposed  beam prediction method in terms of  accuracy and few-shot generalization.

\section{System Model and Problem Formulation}

\subsection{System Model}

\begin{figure}[htbp]
\centerline{\includegraphics[height=0.24\textheight, width=0.85\linewidth]{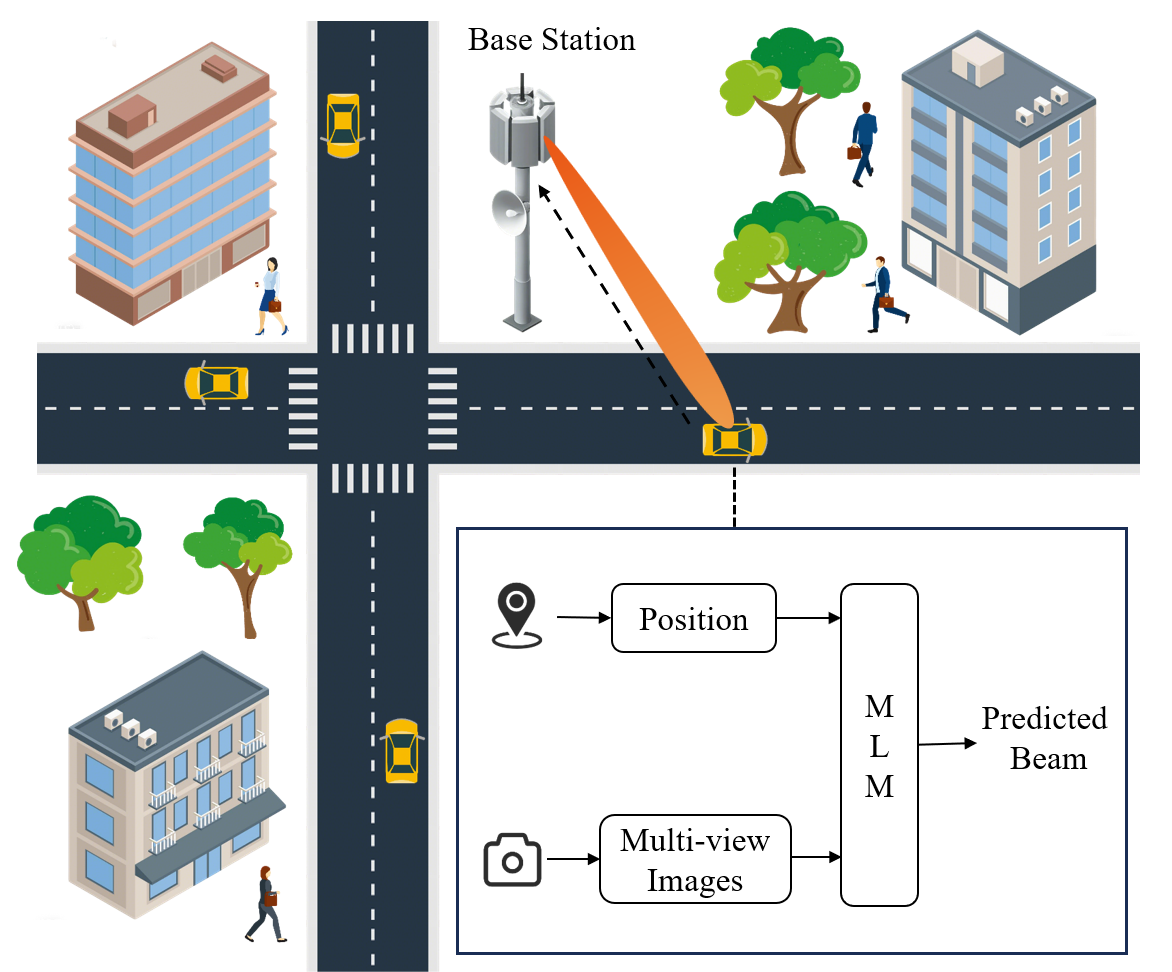}}
\caption{The system model in V2I communication scenario.}
\label{fig:fig1}
\end{figure}

Consider a downlink communication system consisting of a base station (BS) and a mobile station (MS). 
As shown in \Cref{fig:fig1}, the mobile user is equipped with sensors including multiple RGB cameras and a GPS receiver. 
The BS is equipped with a uniform planar array (UPA) of $\textit{N}_{b}$ antennas to communicate with a single-antenna mobile user. 
The communication system employs orthogonal frequency division multiplexing (OFDM) with $\textit{N}_{s}$ subcarriers for information transmission. 
The downlink channel on the $\textit{k}$th subcarrier can be given by
\begin{equation}
\mathbf{h}[k]=\sum_{l=1}^{L} \alpha_{l} e^{-j2\pi f_{k}\tau _{l} + j\mathit{\psi } _{l}} \mathbf{a}\left ( \theta _{l} , \phi_{l}  \right ),
\end{equation}
where \( \alpha_l \), \( \tau_l \), and \( \mathit{\psi}_l \) are the attenuation,  the time delay, and the phase shift of the \( l \)th path respectively, while  \( f_k \) is the frequency of the \( k \)th subcarrier. \( L \) is the total number of multipath components. Additionally, \( \theta_l \) and \( \phi_l \) represent the azimuth and elevation angle of departure for the \( l \)th path respectively, and \( \mathbf{a}(\theta_l, \phi_l) \) is the steering vector of the antenna array at the transmitter. 
When the antenna spacing is set to half a wavelength, the mathematical expression of \( \mathbf{a}(\theta_l, \phi_l) \) can be written as
\begin{equation}
\begin{aligned}
\mathbf{a}(\theta_l, \phi_l) = &  \frac{1}{\sqrt{N_{b}^{h}N_{b}^{v}} } \left[ 1, \dots, e^{j\pi\left[ h \cos \left( \phi_l \right) + v \sin \left( \theta_l \right) \sin \left( \phi_l \right) \right]}, \right. \\
& \left. \dots, e^{j\pi\left[ \left(N_{b}^{h}-1\right) \cos \left( \phi_l \right) +  \left(N_{b}^{v}-1\right) \sin \left( \theta_l \right) \sin \left( \phi_l \right) \right]} \right],
\end{aligned}
\end{equation}
where \( N_b = N_b^h \times N_b^v \), while \( N_b^h \) and \( N_b^v \) represent the number of antenna elements along the horizontal and vertical directions, respectively.

We adopt a pre-defined discrete Fourier transform (DFT) codebook \(\boldsymbol{\mathcal{F}} =\left \{ \mathbf{f}_{m} \right \} _{m=1}^{M} \), where \( M \) represents the size of the entire beamforming codebook set, and \( \mathbf{f}_{m}\in \mathbb{C} ^{N_b\times 1 } \) is the transmission beamforming vector. 
If the BS uses \( \mathbf{f}_{m} \) as the beamforming vector, the received signal on the $\textit{k}$th subcarrier can be expressed as
\begin{equation}
y \left [ k \right ] =\mathbf{h} ^{\mathit{T} } \left [ k \right ] \mathbf{f}_{m}x\left [ k \right ] +n\left [ k \right ],
\end{equation}
where \( x[k] \in \mathbb{C} \) is the transmit signal  and \( n[k] \sim \mathcal{CN}(0, \sigma^2) \) is the additive white Gaussian noise.
\subsection{Problem Formulation}

In this paper, our primary task is to predict the optimal beam index by leveraging the positional information and  multi-view images acquired from the mobile vehicle. 
The aim of beam prediction is to select the optimal beamforming vector \(\mathbf{f} ^{\ast }\) from a set of candidate beams in the codebook \(\boldsymbol{\mathcal{F}}\) to maximize the received signal power. 
Hence, the mathematical formulation of this problem can be defined as
\begin{equation}
\mathbf{f} ^{\ast } =\underset{\mathbf{f}\in \boldsymbol{\mathcal{F}}}{argmax}\frac{1}{\textit{N}_{s}} \sum_{k=1}^{\textit{N}_{s}} \left \|  \mathbf{h}^{\mathit{T} } \left [ k \right ] \mathbf{f}  \right \| ^{2}. 
\end{equation}

We aim to design an artificial intelligence model for predicting the optimal beam index. 
The objective of this model is to learn a mapping function \(\mathbf{\Psi } _{\omega} \) that establishes the relationship between the input data and the optimal beam index. 
The input data \(\mathbf{S} \) consists of multi-view images \(\mathbf{I}_{1}  ,\mathbf{I}_{2} ,\dots ,\mathbf{I}_{N_{c} }\in \mathbb{R} ^{W\times H\times 3}  \) and location information \(\mathbf{g}\in \mathbb{R} ^{3}\), where \(N_{c}\), \(W\), and \(H\) represent the number of cameras mounted on the mobile vehicle, the width and the height of the image, respectively. 
The output of the model is the probability distribution 
\(\mathbf{P}\in \mathbb{R} ^{M\times 1}\) over all beams in the codebook, where the index corresponding to the element with the highest probability represents the optimal beam. 
The mapping function can be mathematically represented as
\begin{equation}
\mathbf{\Psi } _{\omega}: \left \{ \mathbf{S} \right \} \rightarrow \left \{  \mathbf{P}\right \}, 
\end{equation}
where \(\omega\) denotes the parameters of this mapping function. 
In this work, we consider fine-tuning a MLM-based neural network to achieve higher beam prediction accuracy and generalization capability. 
Hence, \(\mathbf{\Psi } _{\omega} \) represents the proposed MLM-based  framework.
\begin{figure*}[htbp]
\centerline{\includegraphics[height=0.34\textheight, width=0.94\linewidth]{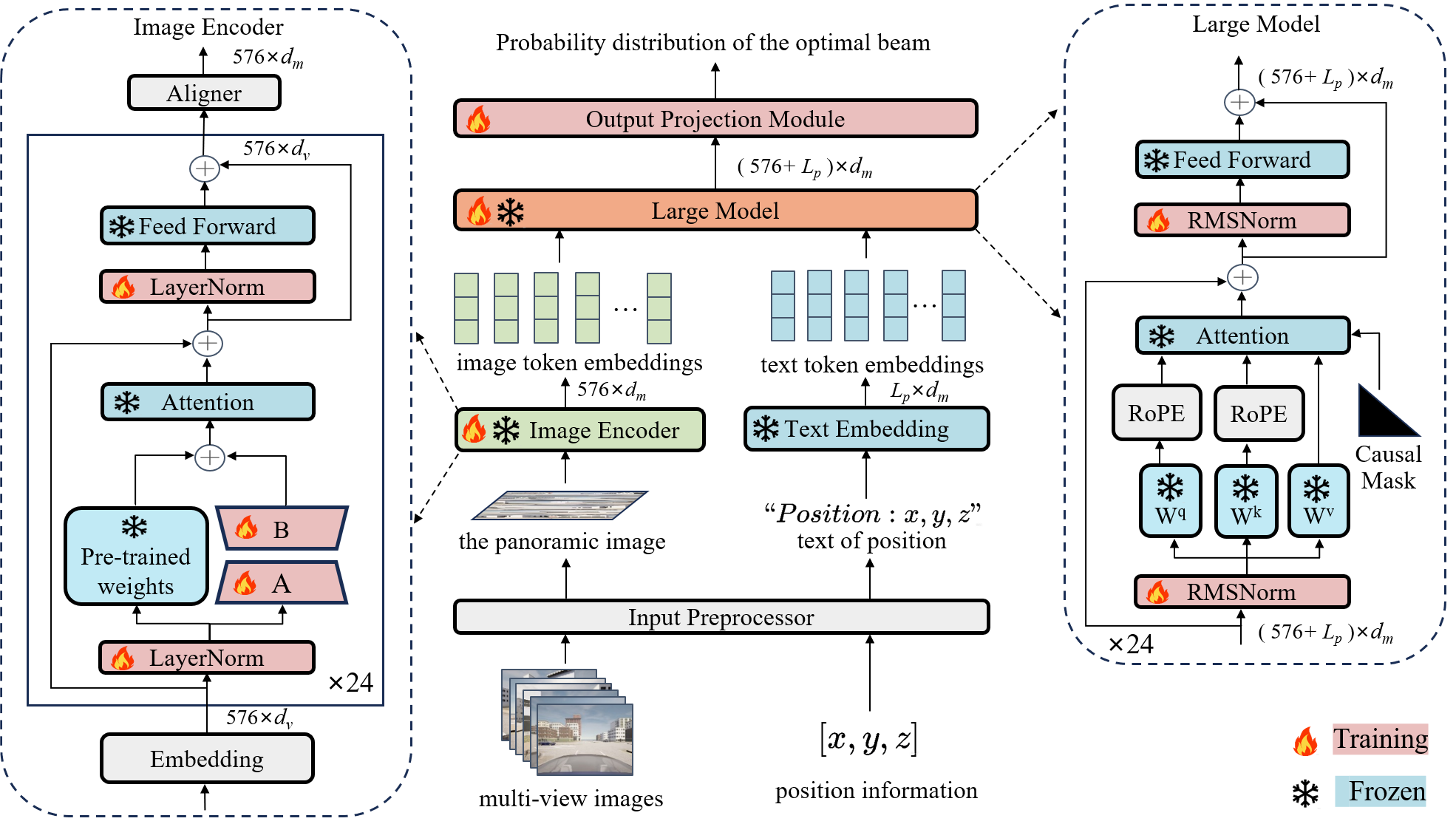}}
\caption{The network architecture of MLM-BP.}
\label{fig:fig2}
\end{figure*}
\section{Multi-Modal Large Models Based Beam Prediction}
In this paper, we propose a MLM-based framework for beam prediction, referred to as MLM-BP, which utilizes position information and multi-view images acquired by the vehicle as inputs. 
The MLM model selected for our study is the Janus-Pro-1B model \cite{ref12} released by DeepSeek. 
Janus-Pro-1B can leverage its multi-modal understanding capability to extract and integrate spatial features from both image and location information. 
As shown in \Cref{fig:fig2}, the framework consists of five modules: input preprocessor, text embedding, image encoder, large model, and output projection. 
The details of the network architecture and the optimization objectives are illustrated below.

\subsection{Network Architecture}
\subsubsection{Input Preprocessor Module}
Given that DeepSeek Janus-Pro-1B requires input images to have a fixed size, \(N_{c}\) multi-view images are resized and concatenated into a single panoramic image \(\mathbf{I}\in \mathbb{R} ^{384\times 384\times 3}  \). 
To improve the convergence rate during model training, we apply mean-standard deviation normalization to obtain the normalized panoramic image \(\mathbf{I}_{n}\). 
For the vehicle's position information, we convert the numerical values into a textual format 
\(\mathbf{T}_{l}\), which is represented as “Position: x, y, z”. 
Finally, we obtain the input data for the model, which includes \(\mathbf{I}_{n}\) and \(\mathbf{T}_{l}\).

\subsubsection{Text Embedding Module}
For the input text \(\mathbf{T}_{l}\), we employ a tokenizer to convert the raw text into a discrete token sequence \(\mathbf{T}_{t}\in \mathbb{R} ^{{L}_{p}\times 1}\), where \({L}_{p}\) represents the sequence length. 
Following tokenization, we map these discrete tokens into continuous high-dimensional representations, resulting in the text token embeddings \(\mathbf{T}_{e}\in \mathbb{R} ^{{L}_{p}\times{d}_{m}} \), where \({d}_{m}\) denotes the feature dimension of the embeddings. 
This embedding helps the subsequent large model understand the spatial information contained in the position for beam prediction.

\subsubsection{Image Encoder Module}
DeepSeek Janus-Pro-1B utilizes the SigLIP \cite{ref13} encoder to extract high-dimensional semantic features from images. 
For the input image \(\mathbf{I}_{n}\), the image encoder performs patch embedding by dividing the image into 576 patches and projecting them into a high-dimensional feature space with positional encoding, resulting in patch embeddings \(\mathbf{P}_{e}\in\mathbb{R} ^{576\times {d}_{v}}\), where \({d}_{v}\) denotes the dimensionality of the embedded features. The embeddings are then fed into encoder blocks for feature extraction. 
Since the vision encoder consists of 24 encoder blocks, to reduce computational cost during training, we apply LoRA  fine-tuning while additionally unfreezing only the layer normalization weights. 
LoRA aims to enable lightweight fine-tuning by simulating full-parameter fine-tuning through the addition of bypass matrices. 
Specifically, we perform LoRA fine-tuning and retraining on the weights of the query, key, and value matrices in the multi-head attention module. 
Assuming the  pre-trained weight matrix is \(\mathbf{W}_{0}\in \mathbb{R} ^{{d}_{v}\times{d}_{v}} \), the LoRA fine-tuning process can be mathematically represented as
\begin{equation}
\mathbf{W} = \mathbf{W} _{0} +\frac{\alpha }{r} \mathbf{B} \mathbf{A}, 
\end{equation}
where \(\mathbf{B}\in \mathbb{R} ^{{d}_{v}\times r} \) and  \(\mathbf{A}\in \mathbb {R} ^{r\times {d}_{v}} \)  contain trainable parameters, while \(r\) and \(\alpha\) represent the rank and scaling factor respectively. 
Generally, \(r\ll d_{v} \) and the  matrix \(\mathbf{B}\) is initialized as a zero matrix, while the  matrix \(\mathbf{A}\) initialized using a random Gaussian distribution. 
For the output of the final encoder block, an aligner is used to project its embedding dimension to \({d}_{m}\), resulting in the image token embeddings \(\mathbf{I}_{e}\in \mathbb{R} ^{576\times{d}_{m}} \).

\subsubsection{Large Model Module}
The large model module of DeepSeek Janus-Pro-1B largely follows  the architectural design of LLaMA \cite{ref14}, comprising a total of 24 decoder blocks. 
To reduce computational cost during training, we freeze the weights of all layers except for those in the root mean square layer normalization (RMSNorm). 
The advantage of RMSNorm lies in reducing computational overhead by eliminating mean normalization, while still maintaining effective scaling. 
In the multi-head attention module of the decoder block, the attention computation for the $\textit{i}$th head, given the query matrix \(\mathbf{Q}^{i}\), key matrix \(\mathbf{K}^{i}\), and value matrix \(\mathbf{V}^{i}\), can be mathematically expressed as
\begin{equation}
\mathrm{ATTENTION}(\mathbf{Q}^{i}, \mathbf{K}^{i}, \mathbf{V}^{i}) = \mathbf{A} \mathbf{V}^{i},
\end{equation}

\begin{equation}
\mathbf{A} = \mathrm{Softmax}\left( \frac{(\mathbf{Q}_{p}^{i})(\mathbf{K}_{p}^{i})^T}{\sqrt{d_{m}}} + \mathbf{M} \right),
\end{equation}
where $\mathbf{Q}_{p}^{i}$ and $\mathbf{K}_{p}^{i}$ are obtained by applying the Rotary Positional Embedding (RoPE), which rotates each input vector according to its position in the input sequence.
$\mathbf{M}$ represents the causal mask, which is used to prevent attending to future tokens by masking out the upper triangular portion of the attention matrix. 
Furthermore, the feed-forward network in the large model employs the SwiGLU activation function, which combines the smoothness of Swish with the gating mechanism of GLU. This design provides an effective activation function that enhances the training of complex and efficient models. 
Finally, the output of the last decoder layer is normalized to obtain the final output of the entire large model, denoted as \(\mathbf{L}_{o}\in \mathbb {R} ^{(576 + {L}_{p})\times {d}_{m}} \).

\subsubsection{Output Projection Module}
To project the high-dimensional output of the large model into a probability distribution over all beams in the codebook, we reduce  \(\mathbf{L}_{o}\) to the matrix \(\mathbf{L}_{m}\in \mathbb {R} ^{{d}_{m}\times 1} \) by computing the mean across the sequence length. Subsequently, the matrix \(\mathbf{L}_{m} \) is passed through a multi-layer perceptron (MLP) with three hidden layers to obtain the final output of the entire framework \(\mathbf{P}\), which can be mathematically formulated as
\begin{equation}
\mathbf{P} = \mathrm{Softmax}\left( \mathrm{MLP}\left (\mathbf{L}_{m}\right )\right).
\end{equation}

\subsection{Optimization objectives}
During the model training process, the proposed neural network predicts the optimal beam probability distribution as \(\mathbf{\widehat{P} } \), while the ground truth \(\mathbf{P}\) is available. Cross-entropy can be used as a loss function to minimize the discrepancy between the predicted beam index and the ground truth. The loss function can be expressed as
\begin{equation}
\mathcal{L} _{B} =- \frac{1}{M} \sum_{i=1}^{M} \mathbf{P} _{i} \log({\mathbf{\widehat{P}}_{i} }). 
\end{equation}

To  evaluate the model's performance, the $\textit{K}$ accuracy  is employed to measure the probability that the correct optimal beam index appears within the top $\textit{K}$ highest predicted probabilities. 
Mathematically, the Top-$\textit{K}$ accuracy is defined as
\begin{equation}
\mathcal{T} _{B} =\frac{1}{D} \sum_{i=1}^{D} \mathcal{I}(\mathrm{\textit{Top-1}}(\mathbf{P}_{i} )\in \mathrm{\textit{Top-K}}(\mathbf{\widehat{P}}_{i} )),
\end{equation}
where $\textit{D}$ is the total number of samples, \(\mathcal{I}\left ( \cdot  \right ) \) is an indicator function that returns 1 if the true beam index is within the top K highest predicted probabilities and 0 otherwise. The function \(\mathrm{\textit{Top-K}}\) represents the set of beam indices corresponding to the top \textit{K} largest values in the predicted probability distribution.

\begin{figure}[htbp]
\centering
\begin{subfigure}[b]{1\linewidth}
    \centering
    \includegraphics[height=0.2\textheight]{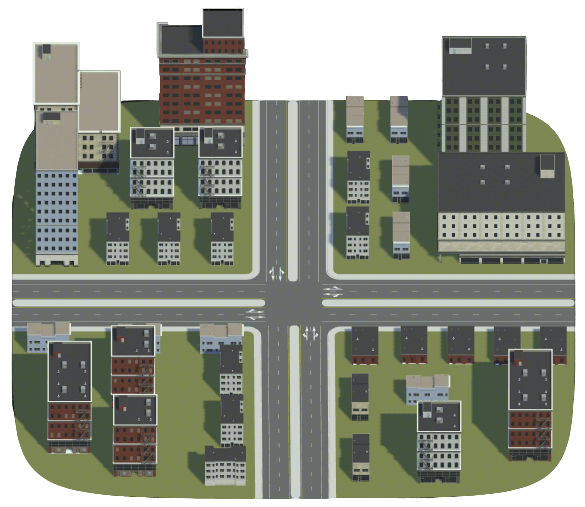}
    \caption{}
    \label{fig:fig3a}
\end{subfigure}

\vspace{0.25cm}

\begin{subfigure}[b]{1\linewidth}
    \centering
    \includegraphics[height=0.15\textheight]{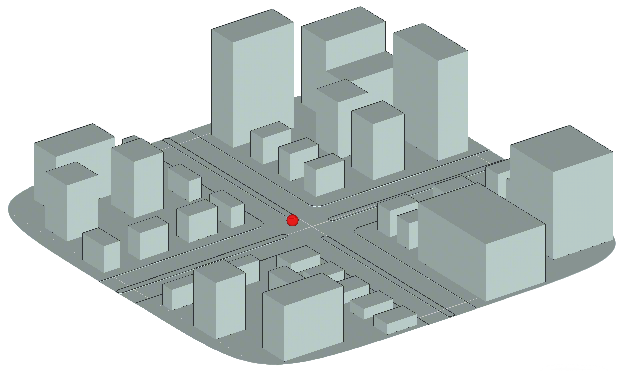}
    \caption{}
    \label{fig:fig3b}
\end{subfigure}

\caption{The environment and channel simulation scenarios. (a) The constructed urban scenario. (b) The ray-tracing scenario.}
\label{fig:fig3}
\end{figure}
\begin{table}[!t]
    \centering
    \caption{Ray Tracing Parameters}
    \label{tab:raytracing}
    \newlength{\colwidth}
    \setlength{\colwidth}{0.235\textwidth - 2\tabcolsep}
    \begin{tabular}{|>{\centering\arraybackslash}p{\colwidth}|>{\centering\arraybackslash}p{\colwidth}|}
        \hline
        Parameter & Value \\
        \hline
        Carrier Frequency & 28 GHz \\
        \hline
         Propagation Model & X3D \\
        \hline
         BS Antennas & 64 \\
        \hline
        Antenna Interval & 0.5 wave-length \\
        \hline
        OFDM Subcarriers & 512 \\
        \hline
        Reflection Order & 6 \\
        \hline
        Diffraction Order & 1 \\
        \hline
         Paths Per Receiver Point & 25 \\
        \hline
    \end{tabular}
\end{table}

\section{Simulation Setup and Performance Analysis}
\subsection{Simulation Setup}
\subsubsection{Dataset}
As illustrated in \Cref{fig:fig3}, following a similar approach to the BUPTCMCC-DataAI-6G dataset \cite{ref15}, we construct an urban scenario with a width of \(W_{s}\) and a length of \(L_{s}\) in the autonomous driving simulator CARLA\cite{ref16}, where the BS is deployed at the roadside. 
The scenario comprises four building complexes and eight roads, with three vehicles serving as mobile stations by traveling along different roads at varying speeds. As the vehicles move along the roads, the onboard cameras capture multi-view images at a time interval of \(T_{s}\) while simultaneously recording the corresponding vehicle positions. 
To simplify the simulation, we import the entire scenario into Blender and convert the buildings and vehicles into cubes of the same size. Then, the simplified scenario is synchronized to the ray tracing programme \cite{ref17} for wireless channel simulation. Through the above steps, a dataset containing MS positions, multi-view images, and the corresponding channel information is constructed. 
Specifically, we set \( L_s =\) 200 \(m\), \( W_s =\) 200 \(m\), and \( T_s =\) 0.02 \(s\). The UPA of the BS is fixed at a height of 6 meters above the ground, while the single antenna of the MS is mounted on top of the vehicle at a height of 1.5 meters. Each vehicle is equipped with \(N_{c} =\) 6 cameras, covering six directions to ensure full 360-degree perception, each capturing images with a resolution of \( W \times H =\) 800 \(\times\) 600 pixels. The parameters for ray tracing are shown in Table~\ref{tab:raytracing}; 
The number of antennas at the base station is \( N_c =\) 64, with \( N_b^h = N_b^v =\) 8. The number of subcarriers in the OFDM system is set to \( N_c =\) 512. 
The pre-defined codebook contains \(M =\) 64 candidate beams, and the optimal beam index is determined by calculating the maximum received power based on the channel information. 
After converting all the channel information into optimal beam indices, the entire dataset is divided into training, validation, and test sets in a ratio of 7:1:2. 
In total, the training set includes 3971 sample points, the validation set includes 567 sample points, and the test set includes 1135 sample points.

\subsubsection{Network and Training Parameters}
During the training process of the proposed MLM-BP framework, we employ LoRA to fine-tune the model. 
The LoRA parameters are set with a rank of \(r\) = 8 and a scaling factor of \(\alpha\) = 32, specifically applied to the query, key, and value matrices within the multi-head attention module of the image encoder. 
The embedding dimensions of the image encoder and the large model are set to \({d}_{v}\) = 1024 and \({d}_{m}\) = 2048, respectively. 
The model is trained using the Adam optimizer with a batch size of 10 and a learning rate of 0.0001 for 200 epochs.
\subsubsection{Baselines}
To validate the effectiveness of the proposed MLM-BP framework, we implement three deep learning-based beam prediction models with a small number of parameters as benchmarks, including a DNN-based model utilizing location, a CNN-based model utilizing multi-view images, and the multi-modal model proposed in \cite{ref8}.

\textbullet \textbf{DNN with position}: It is a DNN-based model that only depends on position. The model consists of 5 linear layers, each containing normalization and residual connections.

\textbullet \textbf{CNN with vision}: It is a CNN-based model that relies only on images. The model consists of 3 convolutional layers, each of which contains batch normalization, ReLU activation, and dropout for regularization.

\textbullet \textbf{ResNet with vision and position}: It is used in \cite{ref8} to implement beam prediction based on image and position data through two sequential stages. In the first stage, a ResNet-50 model, after stripping off the final classifier layer,  is used to extract a  feature vector. The normalized position data is then concatenated with the feature vector, forming a combined vector. The combined vector is fed into a two-hidden-layer MLP network in the second stage.
\begin{figure}[htbp]
\centerline{\includegraphics[height=0.28\textheight]{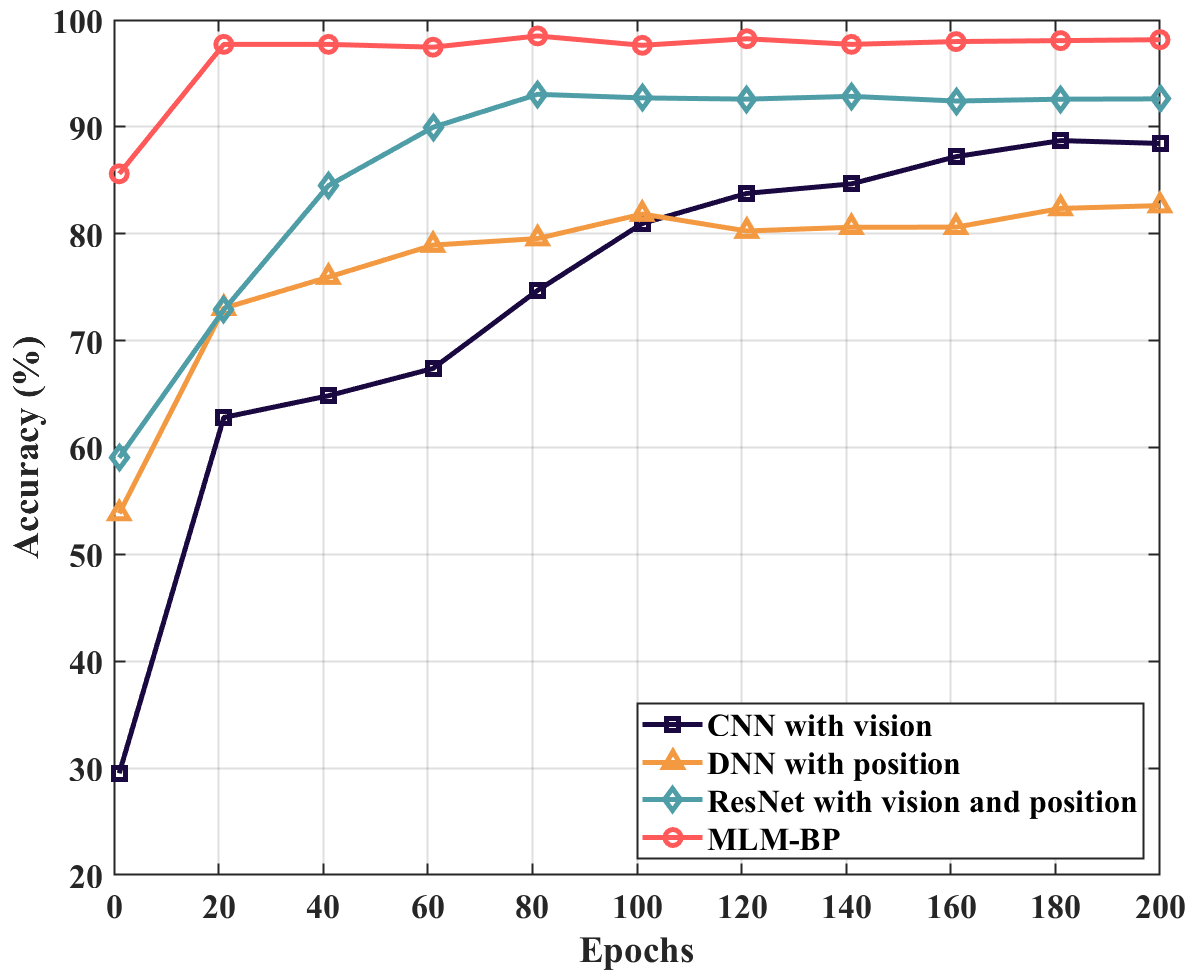}}
\caption{The Top-1 accuracy comparison of MLM-BP and other methods with small parameters on the simulation dataset.}
\label{fig:fig4}
\end{figure}
\subsection{Performance Analysis}
\subsubsection{Performance on the Simulation Dataset}
As illustrated in \Cref{fig:fig4}, the Top-1 accuracy curves of different models exhibit significant differences. The horizontal axis represents the number of training epochs, while the vertical axis indicates the Top-1 accuracy. 
The Top-1 accuracy of MLM-BP rapidly reaches approximately 97.5$\%$ after 20 training epochs and remains stable in subsequent training epochs, ultimately achieving 98.1$\%$. This demonstrates excellent convergence speed and performance stability. 
In contrast, the results reveal that the position-only approach achieved the lowest accuracy, with a Top-1 accuracy of only 82.6$\%$, while the vision-only approach achieved a Top-1 accuracy of 88.4$\%$. Although the vision-position approach integrates both location and image features, achieving a final accuracy of 92.4$\%$, its convergence speed and final accuracy are significantly behind MLM-BP.
The experimental results validate the superior performance of the proposed framework in beam prediction through the effective integration of multi-modal information.

\subsubsection{Few-Shot Prediction}
\begin{figure}[t]
\centerline{\includegraphics[height=0.16\textheight, width=0.87\linewidth]{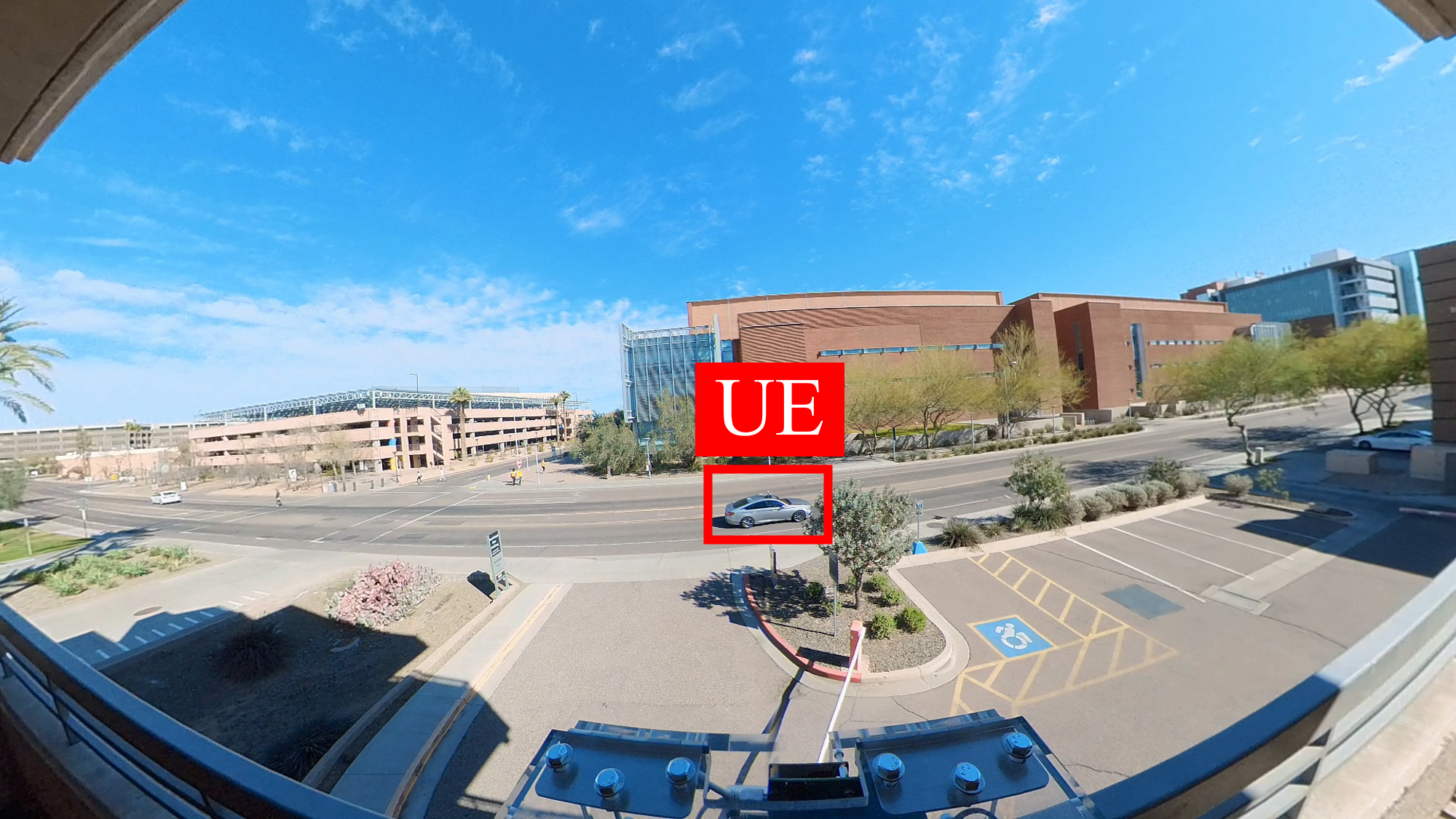}}
\caption{The base station perspective of Scenario 41.}
\label{fig:fig5}
\end{figure}

\begin{figure}[t]
\centering
\begin{subfigure}[b]{1\linewidth}
    \centering
    \includegraphics[height=0.19\textheight]{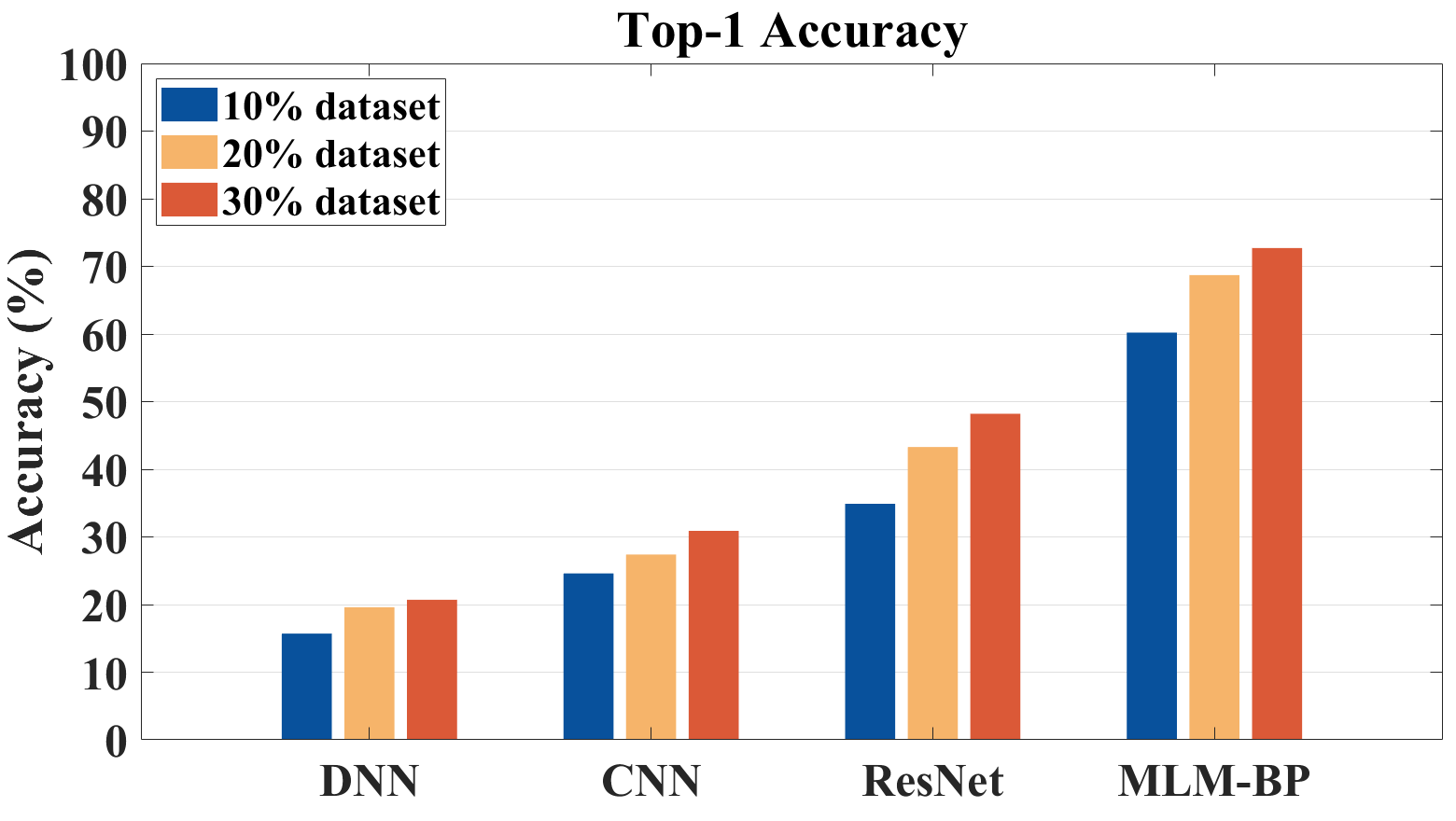}
    \caption{}
    \label{fig:fig6a}
\end{subfigure}

\vspace{0.15cm}

\begin{subfigure}[b]{1\linewidth}
    \centering
    \includegraphics[height=0.19\textheight]{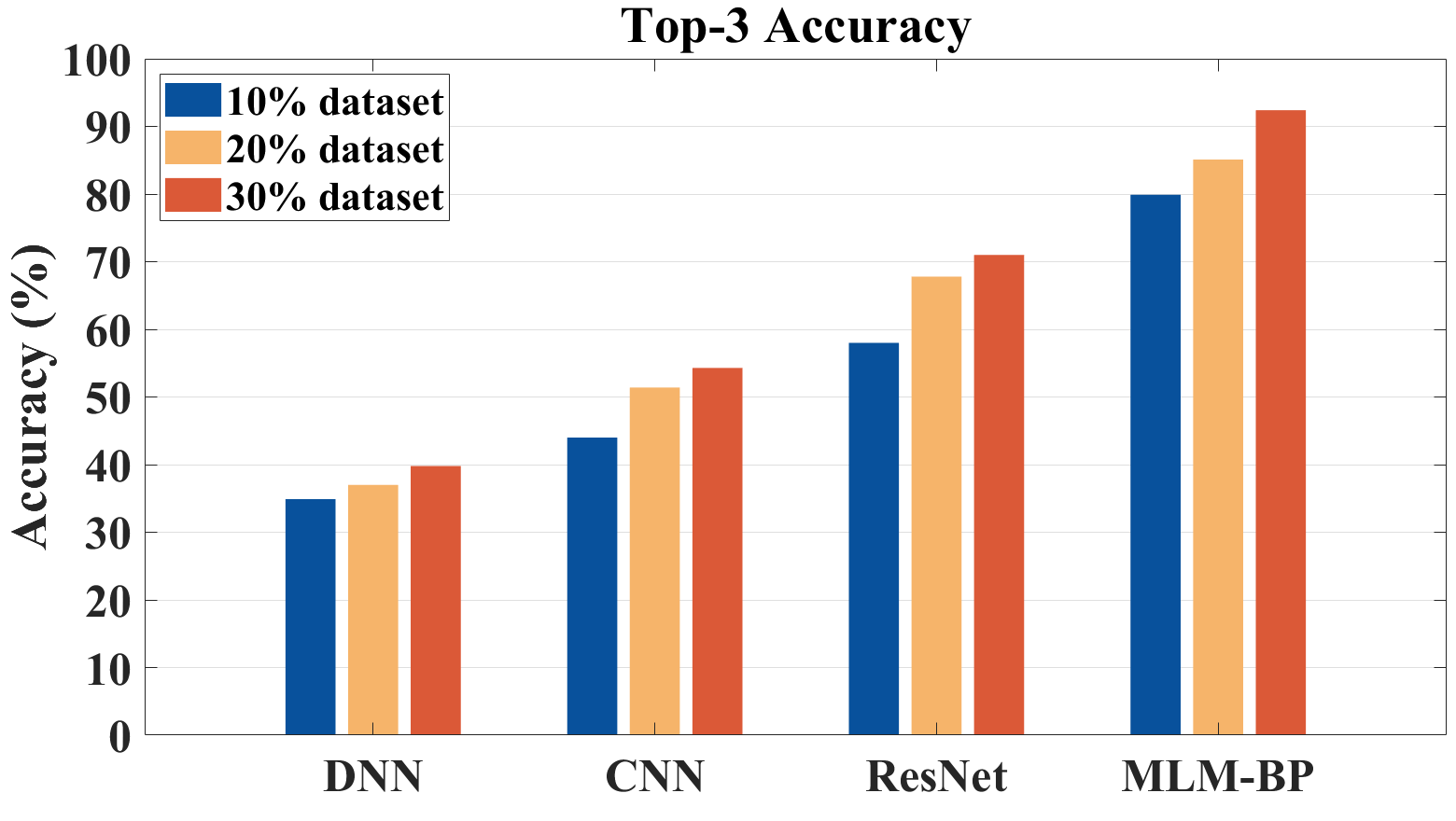}
    \caption{}
    \label{fig:fig6b}
\end{subfigure}

\caption{Few-shot generalization performance on real-world dataset. (a) Top-1 accuracy. (b) Top-3 accuracy.}
\label{fig:fig6}
\end{figure}

The few-shot generalization capability of beam prediction models is crucial for reducing the cost of real-world data collection and network training. 
To evaluate this capability, we assess the proposed model and baselines on the real-world DeepSense dataset \cite{ref18}. Scenario 41 of this dataset, as shown in \Cref{fig:fig5}, contains 22,500 sample points, where each sample includes GPS location information and a single-view image collected by the vehicle, as well as the optimal beam index associated with three uniform linear arrays installed at the base station. The carrier center frequency used is 60 GHz, and each phased array at the base station employs an over-sampled codebook consisting of 64 pre-defined beams. To ensure alignment, the original 192 beams are downsampled to 64 beams, without affecting the total beam coverage area. We randomly select 0.1, 0.2, and 0.3 of the dataset to evaluate the few-shot generalization ability of the four models, and use Top-1 accuracy and Top-3 accuracy as evaluation indicators.
As  presented in \Cref{fig:fig6}, the Top-1 and Top-3 accuracy performances of various models under different training sample ratios are systematically demonstrated. As the  sample ratio increases from 0.1 to 0.3, the Top-1 accuracy of MLM-BP improves from 60.2$\%$ to 72.7$\%$, while the corresponding accuracies of the other three models remain below 50$\%$. Notably, when the sample ratio reaches 0.3, the proposed model achieves a Top-3 accuracy of 92.4$\%$, whereas the Top-3 accuracies of all other models remain below 75$\%$. 
These numerical results objectively verify that the proposed model significantly reduces reliance on large-scale training samples, showcasing strong few-shot generalization capability and practical value in lowering real-world data collection and network training costs.

\section{Conclusion}
In this paper, we  propose the MLM-BP framework, a multi-modal large  model based  beam prediction scheme. The framework fine-tunes the DeepSeek Janus-Pro-1B model to predict the optimal beam index by integrating position and multi-view images. Simulation results in a scenario built in the digital world show that MLM-BP achieves a Top-1 accuracy of 98.1$\%$. Few-shot evaluations on the real-world dataset further demonstrate its strong generalization capability, achieving a Top-3 accuracy of 92.4$\%$ with 30$\%$ of the dataset, while other models remain below 75$\%$. These results highlight the potential of MLM-BP in reducing data collection costs while maintaining high prediction accuracy, making it a promising solution for dynamic vehicular environments.

\section*{Acknowledgment}
This work is supported by  the National Natural Science Foundation of China (No. 62401084), the National Key R\&D Program of China (Grant No. 2023YFB2904805), the BUPT-SICE Excellent Student Creative Foundation (yccx-2004-014), and BUPT-CMCC Joint Innovation Center.

\bibliographystyle{IEEEtran}
\bibliography{IEEEabrv,ref}
\end{document}